\documentclass[a4paper,aps,prd,10pt,preprintnumbers,showpacs,showkeys,twocolumn,superscriptaddress,nofootinbib]{revtex4-1}
\usepackage{orcidlink,graphicx,cmap,amsmath,amsfonts}
\usepackage[utf8]{inputenc}
\usepackage[T1]{fontenc}
\usepackage{booktabs}
\usepackage{hyperref}

\begin{document}

\title{A First-Order Eikonal Framework for Quasinormal Modes, Shadows, Strong Lensing, and Grey-Body Factors in a Scalarized Black-Hole Metric}

\author{Bekir Can Lütfüoğlu}
\email{bekir.lutfuoglu@uhk.cz}
\affiliation{Department of Physics, Faculty of Science, University of Hradec Králové, Rokitanského 62/26, 500 03 Hradec Králové, Czech Republic}

\author{Javlon~Rayimbaev} 
\email{javlon@astrin.uz}
\affiliation{Institute of Theoretical Physics, National University of Uzbekistan, Tashkent 100174, Uzbekistan}
\affiliation{University of Tashkent for Applied Sciences, Gavhar Str. 1, Tashkent 700127, Uzbekistan}
\affiliation{Tashkent State Technical University, Tashkent 100095,
Uzbekistan}


\author{Sardor~Murodov} 
\email{mursardor@ifar.uz}
\affiliation{New Uzbekistan University, Movarounnahr Str. 1, Tashkent 100000, Uzbekistan}
\affiliation{Tashkent International University of Education, Imom Bukhoriy 6, Tashkent 100207, Uzbekistan}

\author{Jakhongir~Kurbanov} \email{jaxongir0903@gmail.com} \affiliation{Kimyo International University in Tashkent, Shota Rustaveli Street 156, Tashkent 100121, Uzbekistan}

\author{Muhammad~Matyoqubov} 
\email{m_matyoqubov@mamunedu.uz} \affiliation{Mamun University, Bolkhovuz Street 2, Khiva 220900, Uzbekistan}

\begin{abstract}
We construct an analytic geodesic-optics description of quasinormal ringing, black-hole shadows, strong lensing, and grey-body factors for the static spherical metric introduced in [Bakopoulos, et. al. arXiv:2310.11919]. Working in a weak-hair regime, we derive closed first-order formulas for the photon-sphere radius, orbital frequency, and Lyapunov exponent. These invariants are then employed within the Schutz--Will WKB approach to obtain eikonal quasinormal frequencies, mapped to shadow and strong-deflection observables through exact identities for static spherical geometries, and used to build a closed analytic form for the transmission probability. At leading eikonal order, these relations are controlled by null geodesics and are therefore spin-universal for test scalar/electromagnetic/gravitational sectors, up to subleading corrections. Besides the standard ringdown--shadow correspondence, we present three additional results: (i) an explicit quality-factor correction, (ii) limiting core-size expansions that show when damping ratios are nearly insensitive to the scalarized core, and (iii) a comparative study of grey-body factors for moderate multipoles and several core-size ratios. The resulting construction provides a concise one-parameter connection from the metric function to ringdown, lensing, and scattering observables.
\end{abstract}

\maketitle

\section{Introduction}

Regularized or near-regular black-hole geometries generated by modified-gravity sectors and scalar fields provide useful laboratories for testing how near-horizon structure affects observables. In this context, the scalarized family discussed in Ref.~\cite{BakopoulosEtAl2024} is particularly attractive because it is asymptotically flat, analytically tractable, and controlled by a small set of physically transparent parameters. More broadly, such frameworks were developed to address persistent limitations of the classical GR picture in extreme-curvature regimes: in particular, the appearance of spacetime singularities (where classical predictivity breaks down) and the need for theoretically consistent effective extensions that can encode high-energy corrections without spoiling well-tested weak-field behavior. Scalarized beyond-Horndeski/DHOST models are especially relevant in this context because they provide a controlled way to parameterize non-Schwarzschild strong-gravity effects, while remaining sufficiently constrained to be confronted with horizon-scale observables such as ringdown, shadow, and lensing data; see, for example, Refs.~\cite{BakopoulosEtAl2024,BakopoulosChatzifotisNakas2024,BakopoulosChatzifotisKarakasis2024,EricesFathi2025}.

Recently, primary scalar hair in shift-symmetric (and parity-preserving) Beyond-Horndeski sectors has been extensively discussed and developed. The baseline black-hole construction is given in Ref.~\cite{BakopoulosEtAl2024}, followed by broader solution-generating analyses in Refs.~\cite{BakopoulosChatzifotisNakas2024,BaakeCisternaHassaineHernandezVera2023}. Thermodynamic development was pushed further in Ref.~\cite{BakopoulosChatzifotisKarakasis2024}, while stability and shadow-oriented phenomenology were explored in Ref.~\cite{EricesFathi2025}. Very recent perturbative progress for this primary-hair class has also appeared in Ref.~\cite{CharmousisIteanuLangloisNoui2025}. Conceptually, these works connect to the earlier shift-symmetric no-hair and hairy-solution literature in Refs.~\cite{HuiNicolis2013,SotiriouZhou2014,BabichevCharmousis2014,BabichevCharmousisLehebel2017,SaravaniSotiriou2019,CreminelliEtAl2020}.

For observational channels, two geometric-optics correspondences are central. First, in the eikonal regime, proper oscillations of black holes, {\it quasinormal frequencies}, \cite{Nollert:1999ji, Kokkotas:1999bd, Bolokhov:2025rng, Konoplya:2011qq, Berti:2009kk} are controlled by the unstable photon orbit \cite{Cardoso2009}. Complementary semianalytic and numerical studies by Malik across several black-hole backgrounds further document eikonal and near-eikonal quasinormal mode (QNM) behavior and related grey-body observables \cite{Malik2023Bumblebee,Malik2024DiracEQG,Malik2024SdSMassive,Malik2025BraneRNdS,Malik2026Dehnen}. Related recent analyses by Dubinsky and collaborators cover asymptotic tails, overtone extraction, asymptotically safe and quantum-corrected backgrounds, and grey-body/eikonal structure in dilatonic and non-minimal Einstein--Yang--Mills black holes \cite{DubinskyZhidenko2024DilatonDS,Dubinsky2024OvertonesTDI,Dubinsky:2024aeu,Dubinsky2025QGCorrectionsSchwarzschild,Dubinsky2025GMGHSGreybody,Dubinsky2025NMEYM}. Second, the same null-geodesic sector determines the black-hole shadow scale and strong-deflection lensing coefficients \cite{Bozza2002,Stefanov2010,PerlickTsupko2022Review}. At the same time, the geodesic/QNM connection is not universally exact for all perturbation sectors and all backgrounds \cite{KonoplyaStuchlik2017,Bolokhov:2023dxq}. This motivates analytic control of the regime where the correspondence is expected to hold. Recent dedicated studies now provide a broad QNM--shadow phenomenology across quantum-corrected, Weyl-inspired, Einstein--Gauss--Bonnet, and scalarized settings, which is closely aligned with the strategy adopted here \cite{Bonanno:2025dry,Konoplya:2025mvj,Konoplya:2024lch,Sekhmani:2026aun,Konoplya:2020bxa,Konoplya:2019xmn,Konoplya:2019fpy,Konoplya:2019goy,Konoplya:2019sns}.  Overall, the recent literature on particle motion, optical phenomena, and the eikonal regime of black holes other compact objects in modified gravity is extensive. Here we mention only a few representative examples \cite{Rahmatov:2026mdo,Asghar:2026que,Hashimoto:2016dfz,Atamurotov:2021imh,Konoplya:2018arm}, referring the reader to the references therein for a more comprehensive overview.

The central motivation of this work is to build a transparent analytic bridge between a concrete scalarized black-hole geometry and multiple observable channels that are often studied separately. Rather than treating ringdown, shadow size, strong lensing, and transmission properties as disconnected outputs, we formulate them within a single parameter-controlled framework tied directly to the scalarized black-hole metric derived in the beyond-Horndeski theory developed in Ref.~\cite{BakopoulosEtAl2024}.

More specifically, the paper is organized around four aims: (i) derive first-order closed expressions for the photon-sphere invariants that control eikonal dynamics, (ii) propagate these invariants consistently to quasinormal frequencies, shadow and lensing observables, and grey-body factors (GBFs), (iii) isolate which combinations of $(\beta,\lambda/M)$ are actually measured by each channel, and (iv) identify the regime where these compact formulas are most useful as controlled approximations for interpretation and parameter inference.

Our broader vision is therefore not only to report formulas, but to provide an interpretable multi-channel map from theory space to observables: a minimal semianalytic scaffold that complements full numerical calculations, clarifies parameter sensitivities, and helps design combined ringdown+shadow+lensing tests of scalarized compact-object models.

The paper is organized as follows. Section~II introduces the geometry and weak-hair bookkeeping. Section~III derives photon-sphere data to first order in $\beta$. Section~IV gives eikonal WKB quasinormal frequencies. Section~V presents the shadow map, damping time, and quality factor. Section~VI derives GBFs in the eikonal limit from the QNM spectrum and provides an illustrative comparison for $\ell=3,4$. Section~VII derives strong-lensing observables and the Stefanov-type correspondence. Section~VIII discusses limiting regimes and parameter degeneracies. Section~IX summarizes.

\section{Geometry and perturbative regime}

The black-hole branch studied here belongs to the shift-symmetric, parity-preserving scalar-tensor sector discussed in Refs.~\cite{BakopoulosEtAl2024,BakopoulosChatzifotisNakas2024}. A convenient compact representation of the underlying theory is the beyond-Horndeski (DHOST-type) action
\small
\begin{widetext}
\begin{equation}
S=\int d^4x\,\sqrt{-g}\,\Bigg[\frac{M_{\rm Pl}^2}{2}R+K(X)+G_4(X)R+G_{4X}(X)\big((\Box\phi)^2-\nabla_\mu\nabla_\nu\phi\,\nabla^\mu\nabla^\nu\phi\big)+\mathcal{L}_{\rm bH}(X,\phi) \Bigg], \label{eq:action-schematic}
\end{equation}
\end{widetext}
\normalsize
where $X\equiv-\tfrac12\nabla_\mu\phi\nabla^\mu\phi$, $G_4(X)$ is the (beyond-)Horndeski coupling function multiplying the Ricci scalar term, while $G_{4X}\equiv dG_4/dX$ controls the associated derivative coupling sector, and shift symmetry is $\phi\to\phi+c$.

We consider the static spherical line element
\begin{equation}
 ds^2=-f(r)dt^2+\frac{dr^2}{f(r)}+r^2\left(d\theta^2+\sin^2\theta\,d\varphi^2\right),
\end{equation}
with metric function taken from Eq.~(9) of Ref.~\cite{BakopoulosEtAl2024}:
\begin{equation}
 f(r)=1-\frac{2M}{r}+\eta q^4\left[\frac{\tfrac{\pi}{2}-\arctan\!\left(\frac{r}{\lambda}\right)}{\frac{r}{\lambda}}+\frac{1}{1+\left(\frac{r}{\lambda}\right)^2}\right]. \label{eq:fexact}
\end{equation}
Here $M$ is the asymptotic mass, $q$ is the scalar-hair integration constant and $\eta$ is the theory coupling.

It is convenient to define
\begin{equation}
\begin{aligned}
\beta &\equiv \eta q^4, \\  x &\equiv \frac{r}{\lambda}, \\
\Phi(x) &\equiv \frac{\pi}{2} - \arctan x, \\
 S(x) &\equiv \frac{\Phi(x)}{x} + \frac{1}{1+x^2},
\end{aligned}
\end{equation}
so that $f(r)=1-2M/r+\beta\,S(r/\lambda)$.

For the static spherical branch and coupling choices of Ref.~\cite{BakopoulosEtAl2024}, Eq.~\eqref{eq:action-schematic} yields the metric function (\ref{eq:fexact}), with phenomenology controlled by three parameters $(M,\beta,\lambda)$. The Schwarzschild limit is restored as $\beta\to0$. In geometric units $G=c=1$, both $M$ and $\lambda$ have dimensions of length, while $\beta$ is dimensionless.

The parameter $\lambda$ is a length scale that controls the radial extent of the scalarized core: it determines how rapidly the modified term decays outside the core and therefore how much of the photon-region dynamics is affected. For $r\gg\lambda$, one finds
\begin{equation}
S\!\left(\frac{r}{\lambda}\right)=\frac{2\lambda^2}{r^2}-\frac{4\lambda^4}{3r^4}+\mathcal{O}(r^{-6}),
\end{equation}
and therefore
\begin{equation}
f(r)=1-\frac{2M}{r}+\frac{2\beta\lambda^2}{r^2}+\mathcal{O}(r^{-4}),
\label{eq:farRNlike}
\end{equation}
which is Reissner--Nordstr\"om-like at leading asymptotic order.

We will work in a weak-hair expansion
\begin{equation}
|\beta|\ll1,
\end{equation}
with the practical consistency condition $|\beta S(3M/\lambda)|\ll1$ near the photon sphere.
This hierarchy is physically motivated. 

The combination $\beta=\eta q^4$ is an effective (dimensionless) hair-coupling amplitude multiplying the full non-Schwarzschild correction in $f(r)$, so a small-$|\beta|$ regime is the natural domain in which the background and all derived observables admit a controlled perturbative expansion. It is also the regime most directly aligned with current strong-gravity phenomenology, where deviations from Schwarzschild are expected to be moderate rather than order unity. By contrast, we do not impose a universal smallness assumption on $\lambda/M$; instead, we keep $y=3M/\lambda$ arbitrary and later analyze both limiting sectors ($y\ll1$ and $y\gg1$), so the only strict perturbative parameter is the coupling amplitude itself. At fixed $\beta$, decreasing $\lambda$ (increasing $y$) suppresses asymptotic corrections as $\sim\lambda^2/r^2$, while increasing $\lambda$ enhances the near-photon-sphere deformation through the growth of $S(y)$ at small $y$.

\section{Photon sphere at first order in \texorpdfstring{$\beta$}{beta}}

Black-hole shadow observables are a direct probe of unstable null geodesics near the photon region and remain central in strong-gravity phenomenology, from early imaging proposals and shape analyses to horizon-scale measurements \cite{EHTM87VI2019,JohannsenPsaltis2010,FalckeMeliaAgol2000,HiokiMaeda2009}. For parametrized rotating geometries, practical shadow-construction formalisms and analytic interpretations were developed in detail in Refs.~\cite{PerlickTsupko2022Review,Konoplya:2021slg}. This motivates deriving the photon-sphere data in closed perturbative form before moving to ringdown and lensing maps. Methodological developments for parametrized compact-object spacetimes and shadow calculations, including axisymmetric reconstruction and wormhole applications, further support an explicit first-order parametrization of photon-sphere observables in combined ringdown--shadow analyses \cite{Younsi:2016azx,Konoplya:2021slg,Bronnikov:2021liv,Churilova:2021tgn,Konoplya:2020xam,Konoplya:2025mvj}.

Additional complementary shadow studies in dispersive media and non-vacuum rotating geometries, together with recent coordinate-independent axisymmetric reconstruction strategies, further emphasize the importance of robust geometric observables in black-hole imaging analyses; see, for example, \cite{Atamurotov:2015nra,Atamurotov:2015xfa,Mirzaev:2025fma}.

For null geodesics, the effective potential is $V_{\rm null}=L_z^2f(r)/r^2$, and the unstable circular orbit satisfies
\begin{equation}
r f'(r)-2f(r)=0.
\label{eq:photoncondition}
\end{equation}
We consider the first-order expansion in terms of $\beta$,
\begin{equation}
r_{\rm ph}=3M+\beta\,\delta r+\mathcal{O}(\beta^2),
\qquad \delta r=\frac{3M}{2}\,\mathcal{F}_r(y),
\end{equation}
where we have introduced
\begin{equation}
\begin{aligned}
y&\equiv\frac{3M}{\lambda},
\\
\mathcal{F}_r(y)&=yS'(y)-2S(y)
=-\frac{3\Phi(y)}{y}-\frac{3+5y^2}{(1+y^2)^2}.
\label{eq:deltar}
\end{aligned}
\end{equation}
This yields the first-order photon-sphere radius
\begin{equation}
r_{\rm ph}=3M\left[1+\frac{\beta}{2}\mathcal{F}_r(y)\right]+\mathcal{O}(\beta^2).
\label{eq:rphresult}
\end{equation}
Because Eq.~\eqref{eq:deltar} gives $\mathcal{F}_r(y)<0$ for $y>0$, a positive coupling ($\beta>0$) shifts the photon sphere inward ($r_{\rm ph}<3M$), while a negative coupling shifts it outward. The magnitude of this displacement is largest for wider cores (smaller $y$).

The photon orbital frequency is
\begin{equation}
\Omega_{\rm ph}=\sqrt{\frac{f(r_{\rm ph})}{r_{\rm ph}^2}}
=\frac{1}{3\sqrt{3}M}\left[1+\frac{3\beta}{2}S(y)\right]+\mathcal{O}(\beta^2),
\label{eq:omegaresult}
\end{equation}
where the shift in $r_{\rm ph}$ cancels out at this order, leaving only $S(y)$.

For the Lyapunov exponent, we use
\begin{equation}
\lambda_{\rm ph}^2=\frac{f(r_{\rm ph})\left[2f(r_{\rm ph})-r_{\rm ph}^2f''(r_{\rm ph})\right]}{2r_{\rm ph}^2},
\end{equation}
which yields
\begin{equation}
\lambda_{\rm ph}=\frac{1}{3\sqrt{3}M}\left[1+\beta\,\mathcal{F}_\lambda(y)\right]+\mathcal{O}(\beta^2),
\label{eq:lambdaresult}
\end{equation}
with
\begin{equation}
\mathcal{F}_\lambda(y)=2S(y)-\frac{y^2}{4}S''(y)
=\frac{3\Phi(y)}{2y}+\frac{3+6y^2-y^4}{2(1+y^2)^3}.
\label{eq:Flambda}
\end{equation}
For $y>0$, both $S(y)$ and $\mathcal{F}_\lambda(y)$ are positive in the domain considered here, so $\beta>0$ increases both $\Omega_{\rm ph}$ and $\lambda_{\rm ph}$, whereas $\beta<0$ decreases them. Since $R_{\rm sh}=1/\Omega_{\rm ph}$, the shadow radius responds in the opposite direction: positive $\beta$ makes the shadow smaller, negative $\beta$ makes it larger.

Equations~\eqref{eq:rphresult}--\eqref{eq:Flambda} encode all geodesic-optics inputs needed for eikonal ringdown, shadows, and strong lensing. In the Schwarzschild limit ($\beta\to0$), they reduce to the standard values
\begin{equation}
r_{\rm ph}=3M,
\qquad
\Omega_{\rm ph}=\lambda_{\rm ph}=\frac{1}{3\sqrt{3}M},
\label{eq:schw-photon-limit}
\end{equation}
in agreement with the classic Schwarzschild shadow/photon-orbit results of Synge and Luminet~\cite{Synge1966,Luminet1979}, later used in modern eikonal analyses~\cite{Cardoso2009,PerlickTsupko2022Review}.

\begin{figure*}[t]
\centering
\includegraphics[width=0.98\linewidth]{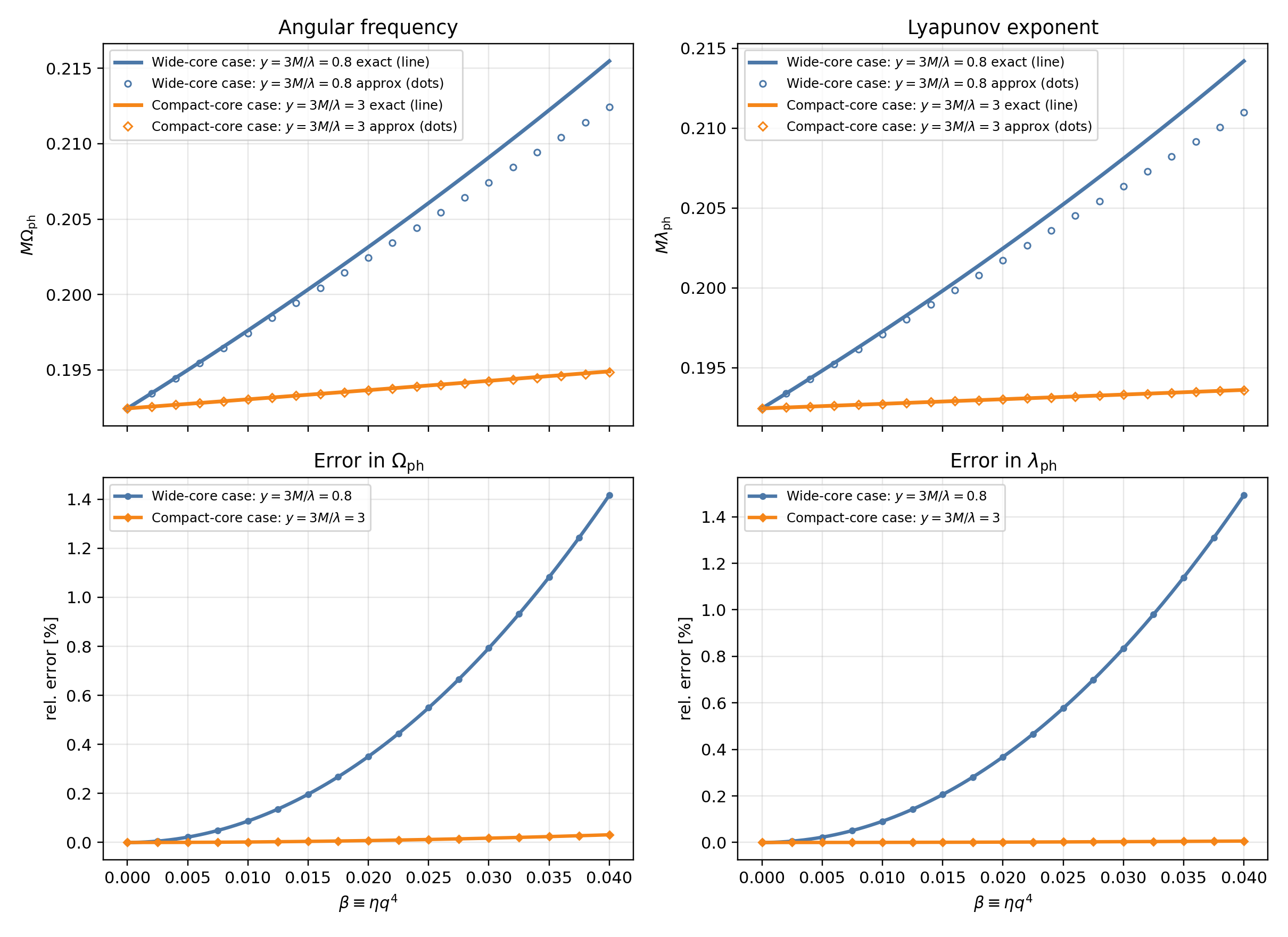}
\caption{Exact numerical data versus first-order analytic approximations for $\Omega_{\rm ph}$ and $\lambda_{\rm ph}$ in the scalarized metric of Eq.~\eqref{eq:fexact}. Top panels: $M\Omega_{\rm ph}$ and $M\lambda_{\rm ph}$ as functions of $\beta$ for $y=3M/\lambda=0.8$ and $y=3$. Bottom panels: corresponding relative errors (in percent). Numerical points are obtained from direct solution of Eq.~\eqref{eq:photoncondition} with the full metric.}
\label{fig:omega-lambda-main2-check}
\end{figure*}

To benchmark Eqs.~\eqref{eq:omegaresult} and \eqref{eq:lambdaresult} against accurate numerical data, we solve Eq.~\eqref{eq:photoncondition} directly with the full metric \eqref{eq:fexact} at each coupling value $\beta$, and then evaluate $\Omega_{\rm ph}$ and $\lambda_{\rm ph}$ from their exact definitions. Figure~\ref{fig:omega-lambda-main2-check} shows this comparison for two representative core-size ratios: a wide-core case $y=3M/\lambda=0.8$ and a compact-core case $y=3$.

Over the interval $0\le \beta\le 0.04$, the first-order approximation remains quantitatively reliable. At $\beta=0.04$, the wide-core case has relative errors
\begin{equation}
\delta\Omega=1.42\%,
\qquad
\delta\lambda=1.49\%,
\end{equation}
while for the compact-core case we find
\begin{equation}
\delta\Omega=0.031\%,
\qquad
\delta\lambda=0.006\%.
\end{equation}
Thus the perturbative formulas are especially accurate when $\lambda\lesssim M$ (large $y$), and remain at the percent level even in the wider-core regime.

\section{Eikonal quasinormal modes from first-order WKB}

The WKB framework remains the standard semianalytic link between barrier-peak data and complex quasinormal frequencies, especially in the eikonal and moderate-overtone domains. Starting from the Schutz--Will/Iyer--Will setup and high-order improvements \cite{SchutzWill1985,IyerWill1987,Konoplya:2026fqh,Konoplya2003WKB}, extensive applications span Schwarzschild--de Sitter, Einstein--Aether, and improved Pad\'e constructions \cite{MatyjasekOpala2017,KonoplyaZhidenko2004SdSHighOvertones,Bolokhov:2024ixe,KonoplyaZhidenkoZinhailo2019,KonoplyaZhidenko2007EAether}. Recent examples include Bolokhov's analyses in higher-curvature and quantum-corrected contexts \cite{Bolokhov2025EffectiveQG,Bolokhov2024Holonomy,Bolokhov2023,Bolokhov2024EulerHeisenberg,Moura:2021eln,Moura:2021nuh,Moura:2022gqm}; Skvortsova's studies of lower-dimensional and near-extreme compact-object backgrounds \cite{Skvortsova2025RegularExtreme,Skvortsova2024QOS,Skvortsova2025QuantumCorrected,Skvortsova2024Spectrum2p1}; L\"utf\"uo\u{g}lu's work in Weyl, Einstein--Yang--Mills, and asymptotically safe settings \cite{Lutfuoglu2025ProperTime,Lutfuoglu2026ASG,Lutfuoglu2025Weyl,Lutfuoglu2025EYM}; and recent time-domain/WKB studies by Dubinsky on charged and quantum-corrected black holes \cite{Dubinsky2024OvertonesTDI,Dubinsky:2024aeu,Dubinsky2025QGCorrectionsSchwarzschild,Dubinsky2025GMGHSGreybody,Dubinsky2025NMEYM}.

For master perturbations,
\begin{equation}
\frac{d^2\Psi}{dr_*^2}+\left[\omega^2-V_\ell(r)\right]\Psi=0,
\qquad
\frac{dr_*}{dr}=\frac{1}{f(r)},
\end{equation}
with quasinormal-mode boundary conditions of purely ingoing waves at the event horizon and purely outgoing waves at spatial infinity,
\begin{equation}
\Psi\propto e^{-i\omega r_*} \quad (r_*\to-\infty),
\qquad
\Psi\propto e^{+i\omega r_*} \quad (r_*\to+\infty).
\end{equation}
Conceptually, the first-order WKB construction used here is based on matching asymptotic WKB series at the two extremities to a local Taylor expansion of the potential near its maximum, thereby reducing the QNM boundary-value problem to a peak-local quantization condition. The relation given in Eq.~\eqref{eq:SW2} should therefore be viewed as the leading Schutz--Will quantization condition, which in many recent black-hole perturbation analyses is used either directly at first order or as the baseline for higher-order WKB and WKB--Pad\'e refinements \cite{Lutfuoglu:2025kqp,Lutfuoglu:2026uzy,Konoplya:2023ahd,Bolokhov:2025fto,Albuquerque:2023lhm,Eniceicu:2019npi,Barrau:2019swg,Konoplya:2005sy,Zinhailo:2019rwd,Chen:2019dip,Skvortsova:2023zca,Konoplya:2006gq,Bronnikov:2019sbx,Dubinsky:2024aeu,Churilova:2021tgn,Konoplya:2001ji,Momennia:2018hsm,Zhidenko:2003wq,Fernando:2012yw,Dubinsky:2025fwv,Bolokhov:2025lnt,Bolokhov:2025aqy,Skvortsova:2026unq,Skvortsova:2024msa}.
With $L\equiv\ell+\tfrac12$, the first-order Schutz--Will condition is \cite{SchutzWill1985,IyerWill1987}
\begin{equation}
\frac{iQ_0}{\sqrt{2Q_0''}}=n+\frac12,
\qquad
Q(r)=\omega^2-V_\ell(r),
\label{eq:SW2}
\end{equation}
Here $Q_0\equiv Q(r_0)$ and $Q_0''\equiv \frac{d^2Q}{dr_*^2}\Big|_{r_0}$ are evaluated at the peak $r_0$ of the effective potential, and $n=0,1,2,\dots$ denotes the overtone index. In the eikonal limit one obtains
\begin{equation}
\omega_{\ell n}=L\Omega_{\rm ph}-i\left(n+\frac12\right)\lambda_{\rm ph}
+\frac{U(r_{\rm ph})}{2L\Omega_{\rm ph}}+\mathcal{O}(L^{-2}).
\label{eq:wkbmaster2}
\end{equation}
Here $U(r)$ is the spin-dependent subleading potential term (for a test scalar, $U(r)=f(r)f'(r)/r$). In the strict eikonal sector, the leading terms in Eq.~\eqref{eq:wkbmaster2} depend only on the null-orbit invariants $(\Omega_{\rm ph},\lambda_{\rm ph})$, so scalar, electromagnetic, and gravitational test fields share the same leading correspondence. Spin dependence re-enters through the $\mathcal{O}(L^{-1})$ correction and through possible non-geometric couplings of the perturbation sector.

Keeping only the leading eikonal piece and substituting Eqs.~\eqref{eq:omegaresult}, \eqref{eq:lambdaresult},
\begin{align}
\omega_{\ell n}
&=\frac{L}{3\sqrt{3}M}\left[1+\frac{3\beta}{2}S(y)\right]
-i\frac{\left(n+\frac12\right)}{3\sqrt{3}M}\left[1+\beta\mathcal{F}_\lambda(y)\right]
\nonumber\\
&\quad +\mathcal{O}(\beta^2,L^{-1}).
\label{eq:wkbexpanded}
\end{align}
Equation~\eqref{eq:wkbexpanded} then reduces to
\begin{equation}
\begin{aligned}
\omega_{\ell n}^{\rm Sch}
&= \frac{1}{3\sqrt{3}M}\left[L - i\left(n+\frac12\right)\right] + \mathcal{O}(L^{-1}), \\
&= \frac{1}{3\sqrt{3}M}\left[\left(\ell+\frac12\right) - i\left(n+\frac12\right)\right] + \mathcal{O}(L^{-1}).
\end{aligned}
\label{eq:mashhoon-limit}
\end{equation}
which is the well-known eikonal Schwarzschild formula first derived by Mashhoon~\cite{Mashhoon1985}.
Thus the scalarized deformation modifies oscillation and damping through distinct response functions $S(y)$ and $\mathcal{F}_\lambda(y)$.
At fixed $(\ell,n,y)$, positive $\beta$ raises $\operatorname{Re}\omega_{\ell n}$ through $S(y)$ and increases $|\operatorname{Im}\omega_{\ell n}|$ through $\mathcal{F}_\lambda(y)$, implying faster oscillations and faster damping. The opposite trend holds for negative $\beta$.

An important caveat should be stated here, which also applies to the subsequent sections. The eikonal expressions and the correspondences employed below—relating null geodesics to spectral characteristics such as QNMs, GBFs, strong-lensing observables, the Lyapunov exponent, and orbital frequencies — are valid only under the assumption that the effective potential is of a WKB-admissible form, i.e., it possesses a single peak and decays monotonically at the boundaries. If the potential develops a double-well structure \cite{Guo:2022ghl,Konoplya:2025hgp}, the eikonal correspondence breaks down. A similar failure may occur in certain higher-curvature theories \cite{Bolokhov:2023dxq,Konoplya:2017wot}, where the effective potential acquires a negative gap that deepens without bound as $\ell \to \infty$, leading to an instability of the background \cite{Takahashi:2010gz,Dotti:2004sh,Konoplya:2017zwo,Konoplya:2017ymp,Gleiser:2005ra}. 

Furthermore, in asymptotically de Sitter spacetimes the correspondence between QNMs and null geodesics (and related shadow or lensing properties) may hold only partially, since the additional de Sitter branch of modes cannot be captured by the WKB approximation even in the eikonal limit \cite{Konoplya:2022gjp}. The configuration considered in the present work does not fall into any of these pathological cases; therefore, the correspondence is exact in the eikonal limit and remains a good approximation for small and moderate values of $\ell$.     

\section{Shadow scale, damping time, and quality factor}

For static spherical geometries, the shadow radius $R_{\rm sh}$ for a distant static observer equals the critical impact parameter $b_c$,
\begin{equation}
R_{\rm sh}=b_c=\frac{r_{\rm ph}}{\sqrt{f(r_{\rm ph})}}=\frac{1}{\Omega_{\rm ph}}.
\label{eq:shadow2def}
\end{equation}
Using Eq.~\eqref{eq:omegaresult},
\begin{equation}
R_{\rm sh}=3\sqrt{3}M\left[1-\frac{3\beta}{2}S(y)\right]+\mathcal{O}(\beta^2).
\label{eq:shadow2}
\end{equation}
Equation~\eqref{eq:shadow2} makes the coupling trend explicit: for $\beta>0$ the fractional correction is negative, so the apparent shadow size decreases; for $\beta<0$ it increases. The dependence is strongest at small $y$, where $S(y)$ is largest.
Then the leading ringdown-shadow relation takes the standard form
\begin{equation}
\begin{aligned}
\operatorname{Re}\omega_{\ell n}&=\frac{L}{R_{\rm sh}}+\mathcal{O}(L^{-1}),
\\
\operatorname{Im}\omega_{\ell n}&=-\left(n+\frac12\right)\lambda_{\rm ph}+\mathcal{O}(L^{-1}).
\end{aligned}
\end{equation}

A convenient damping-time estimate is
\begin{equation}
\tau_n\equiv\frac{1}{|\operatorname{Im}\omega_{\ell n}|}
=\frac{3\sqrt{3}M}{n+\frac12}\left[1-\beta\mathcal{F}_\lambda(y)\right]+\mathcal{O}(\beta^2).
\label{eq:taun}
\end{equation}

As an additional analytic quantity, define the eikonal quality factor
\begin{equation}
\mathcal{Q}_n\equiv\frac{\operatorname{Re}\omega_{\ell n}}{2|\operatorname{Im}\omega_{\ell n}|}
\simeq \frac{L}{2n+1}\left[1+\beta\,\mathcal{D}(y)\right],
\label{eq:Qfactor}
\end{equation}
where
\begin{equation}
\mathcal{D}(y)=\frac{3}{2}S(y)-\mathcal{F}_\lambda(y)=\frac{2y^4}{(1+y^2)^3}>0.
\label{eq:Dy}
\end{equation}
Inequality~\eqref{eq:Dy} implies that for positive $\beta$, the eikonal quality factor (\ref {eq:Qfactor}) is enhanced for all $y$.
Because $\mathcal{D}(y)=2y^4/(1+y^2)^3$ is positive and bounded, the quality-factor correction remains moderate: it approaches the Schwarzschild value in both extreme limits ($y\ll1$ and $y\gg1$) and the deviation is maximal at intermediate core size ($y=\sqrt{2}$).
In the Schwarzschild limit ($\beta\to0$), the shadow radius, damping time, and eikonal quality factor reduce to their standard Schwarzschild forms:
\begin{equation}
R_{\rm sh}=3\sqrt{3}M,
\quad
\tau_n=\frac{3\sqrt{3}M}{n+\frac12},
\quad
\mathcal{Q}_n=\frac{L}{2n+1},
\label{eq:schw-shadow-quality-limit}
\end{equation}
with the shadow scale traced back to the classic works~\cite{Synge1966,Luminet1979} and the eikonal QNM form to Mashhoon~\cite{Mashhoon1985}.

\section{Grey-body factors from the eikonal QNM correspondence}

For a fixed multipole $\ell$, the GBF, also called the transmission probability, is defined by
\begin{equation}
\Gamma_\ell(\omega)\equiv |T_\ell(\omega)|^2,
\qquad
|R_\ell(\omega)|^2+|T_\ell(\omega)|^2=1,
\label{eq:gbfdef}
\end{equation}
where $R_\ell$ and $T_\ell$ are the reflection and transmission amplitudes in the one-dimensional scattering problem for the master wave equation. The associated scattering boundary conditions are
\begin{equation}
\begin{aligned}
&\Psi\to e^{-i\omega r_*}+R_\ell(\omega)e^{+i\omega r_*} \quad (r_*\to+\infty),
\\
&\Psi\to T_\ell(\omega)e^{-i\omega r_*} \quad (r_*\to-\infty),
\end{aligned}
\end{equation}
which imply flux conservation $|R_\ell|^2+|T_\ell|^2=1$ for real-frequency scattering in this static background.

In the eikonal regime, Ref. \cite{Konoplya:2024lir,Konoplya:2024vuj} gives the QNM--GBF correspondence in terms of the fundamental mode $\omega_{\ell 0}$:
\begin{equation}
\Gamma_\ell(\omega)
=\left[1+\exp\!\left(2\pi\,\frac{\omega^2-\operatorname{Re}(\omega_{\ell 0})^2}{4\operatorname{Re}(\omega_{\ell 0})\operatorname{Im}(\omega_{\ell 0})}\right)\right]^{-1}.
\label{eq:gbf35}
\end{equation}
Since $\operatorname{Im}(\omega_{\ell 0})<0$ for damped modes, it is convenient to rewrite this as
\begin{equation}
\Gamma_\ell(\omega)
=\left[1+\exp\!\left(-\pi\,\frac{R_{\rm sh}}{L\lambda_{\rm ph}}\left[\omega^2-\frac{L^2}{R_{\rm sh}^2}\right]\right)\right]^{-1}
+\mathcal{O}(L^{-1}),
\label{eq:gbflogistic}
\end{equation}
with $L\equiv \ell+\frac12$. Here we used the eikonal identities 
\begin{equation}
    \begin{aligned}
\operatorname{Re}(\omega_{\ell 0})&=L\Omega_{\rm ph}=L/R_{\rm sh}+\mathcal{O}(L^{-1}), \\ \operatorname{Im}(\omega_{\ell 0})&=-\lambda_{\rm ph}/2+\mathcal{O}(L^{-1}).        
    \end{aligned}
\end{equation}
Thus, the GBF has a Fermi-type profile centered at $\omega\simeq L/R_{\rm sh}$, with width governed by the instability scale $\lambda_{\rm ph}$.

Using Eqs.~\eqref{eq:shadow2} and \eqref{eq:lambdaresult},
\begin{equation}
\begin{array}{rcl}
\dfrac{L^2}{R_{\rm sh}^2}&=&\dfrac{L^2}{27M^2}\left[1+3\beta S(y)\right]+\mathcal{O}(\beta^2),
\\
\dfrac{R_{\rm sh}}{L\lambda_{\rm ph}}
&=&\dfrac{27M^2}{L}\left[1+\beta\left(\frac{3}{2}S(y)+\mathcal{F}_\lambda(y)\right)\right]^{-1}
+\mathcal{O}(\beta^2),
\end{array}
\label{eq:omegac2prod}
\end{equation}

Substituting \eqref{eq:omegac2prod} into \eqref{eq:gbflogistic} gives the analytic GBF in our scalarized metric:
\begin{equation}
\begin{aligned}
\Gamma_\ell(\omega)
&=\left[1+\exp\!\left(-\frac{27\pi M^2}{L}\,\frac{\omega^2-\frac{L^2}{27M^2}\left(1+3\beta S(y)\right)}{1+\beta\left(\frac{3}{2}S(y)+\mathcal{F}_\lambda(y)\right)}\right)\right]^{-1}\\
&+\mathcal{O}(\beta^2,L^{-1}).
\end{aligned}
\label{eq:gbfmain}
\end{equation}
Expanding the exponent to first order in $\beta$,
\small
\begin{align}
\Gamma_\ell(\omega)
&=\left[1+\exp\!\left(-\frac{27\pi M^2}{L}\left\{\Delta_0(\omega)+\beta\,\Delta_1(\omega,y)\right\}\right)\right]^{-1} \nonumber \\
&+\mathcal{O}(\beta^2,L^{-1}),
\label{eq:gbfexpanded}
\\
\Delta_0(\omega)&\equiv\omega^2-\frac{L^2}{27M^2},
\\
\Delta_1(\omega,y)&\equiv
-\left(\frac{3}{2}S(y)+\mathcal{F}_\lambda(y)\right)\omega^2
+\frac{L^2}{27M^2}\left(\frac{9}{2}S(y)+\mathcal{F}_\lambda(y)\right).
\end{align} \normalsize

Equations~\eqref{eq:gbfmain}--\eqref{eq:gbfexpanded} provide a closed analytic GBF model directly in terms of $(M,\beta,\lambda)$ through $y=3M/\lambda$, $S(y)$, and $\mathcal{F}_\lambda(y)$. The half-transmission point is controlled by
\begin{equation}
\Gamma_\ell(\omega_*)=\frac12
\quad\Longleftrightarrow\quad
\omega_*\simeq\frac{L}{R_{\rm sh}}
=\frac{L}{3\sqrt{3}M}\left[1+\frac{3\beta}{2}S(y)\right],
\end{equation} 
while the steepness of the transition is set by $L\lambda_{\rm ph}/R_{\rm sh}\propto L$.
Hence the parameter trends are transparent: increasing $\ell$ steepens the transition, increasing $\beta$ shifts the half-transmission point to larger frequency through $S(y)$, and changing $y$ controls both the center and width through $(S,\mathcal{F}_\lambda)$. In particular, for fixed positive $\beta$, the shift weakens as $y$ becomes large because $S(y)\sim2/y^2$.

\begin{figure*}[t]
\centering
\includegraphics[width=0.99\linewidth]{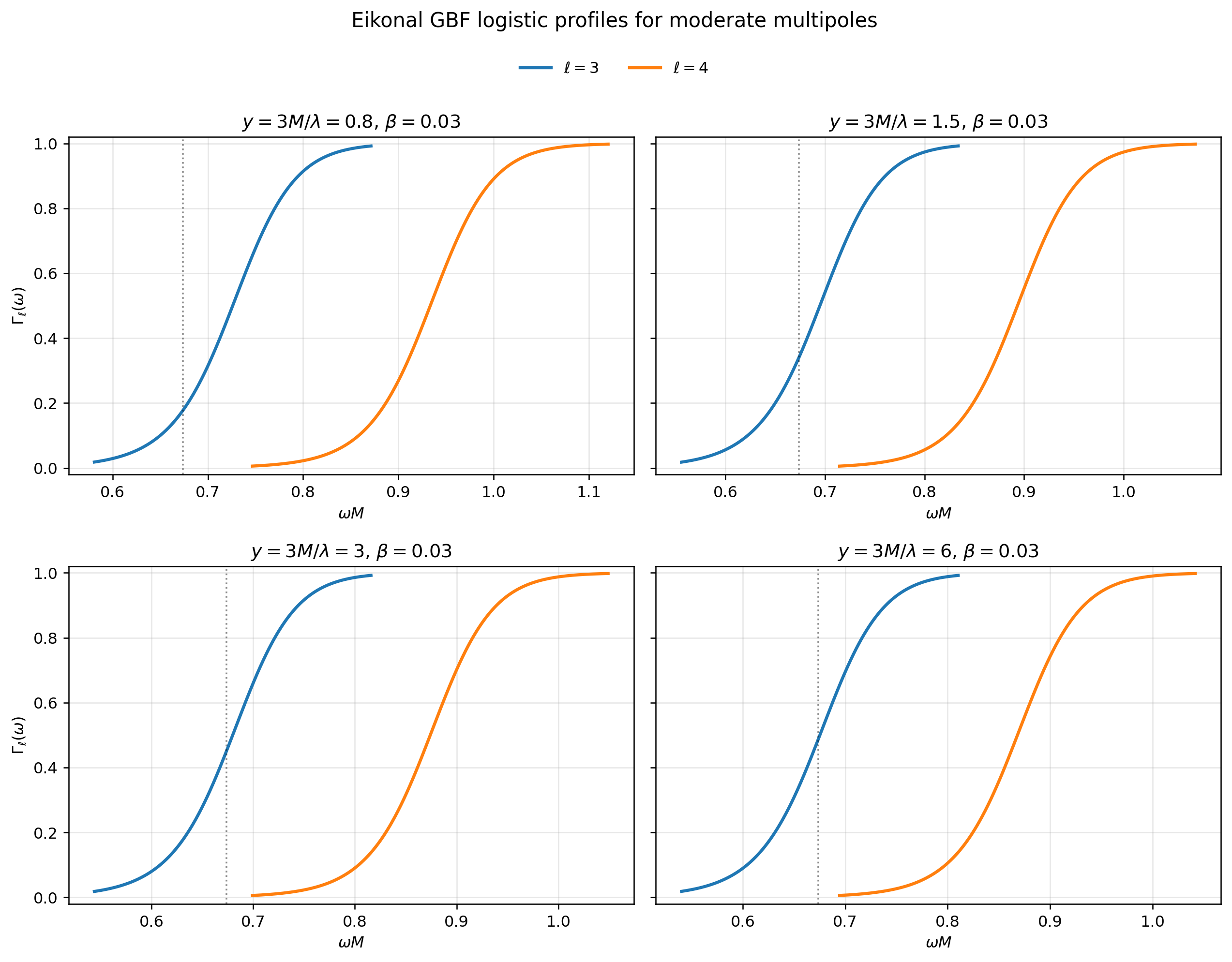}
\caption{Analytic eikonal GBFs $\Gamma_\ell(\omega)$ from Eq.~\eqref{eq:gbfmain} for multiple moderate multipoles, $\ell=3,4$, and several black-hole core-size ratios $y=3M/\lambda=0.8,1.5,3,6$ at fixed coupling $\beta=0.03$ (with $M=1$ units). A larger $\ell$ produces a steeper transition, while varying $y$ shifts the center and width through $S(y)$ and $\mathcal{F}_\lambda(y)$.}
\label{fig:gbf-multiell-params-main2}
\end{figure*}

Figure~\ref{fig:gbf-multiell-params-main2} shows representative GBF profiles for moderate multipoles ($\ell=3,4$) and several core-size ratios at fixed weak-hair coupling. As expected from Eq.~\eqref{eq:gbflogistic}, the transition is visibly steeper for $\ell=4$ than for $\ell=3$. Across the four $y$-values, the center shifts through $R_{\rm sh}^{-1}$ (equivalently $S(y)$), while the transition width tracks $\lambda_{\rm ph}$ through $\mathcal{F}_\lambda(y)$. This confirms that even outside asymptotically large-$\ell$ values, the first-order eikonal approximation captures the qualitative parameter dependence needed for comparative GBF studies.

\section{Strong lensing and the ringdown--lensing map}

The strong-deflection coefficient can be written as \cite{Bozza2002,Stefanov2010}
\begin{equation}
\bar a=\sqrt{\frac{2}{2f(r_{\rm ph})-r_{\rm ph}^2 f''(r_{\rm ph})}},
\end{equation}
where all metric functions are evaluated at the photon-sphere radius $r=r_{\rm ph}$.

As in the general static case,
\begin{equation}
u_m\equiv\frac{r_{\rm ph}}{\sqrt{f(r_{\rm ph})}}=\frac{1}{\Omega_{\rm ph}},
\qquad
\bar a=\frac{\Omega_{\rm ph}}{\lambda_{\rm ph}}.
\label{eq:umabar2}
\end{equation}
Here $u_m$ is the critical impact parameter (minimum impact parameter for capture), i.e. the impact parameter of null rays asymptotically approaching the unstable photon orbit; it coincides with the shadow radius for a distant static observer.
Using Eq.~\eqref{eq:umabar2} together with the first-order geodesic expansions, we obtain
\begin{equation}
\begin{aligned}
u_m&=3\sqrt{3}M\left[1-\frac{3\beta}{2}S(y)\right]+\mathcal{O}(\beta^2),
\\
\bar a&=1+\beta\mathcal{D}(y)+\mathcal{O}(\beta^2),
\end{aligned}
\label{eq:umabarexp}
\end{equation}
with the same $\mathcal{D}(y)$ as in Eq.~\eqref{eq:Dy}.

The observable strong-lensing quantities are
\begin{equation}
\theta_\infty=\frac{u_m}{D_{OL}},
\quad
\mathcal{R}=\exp\!\left(\frac{2\pi}{\bar a}\right),
\quad
r_m=2.5\log_{10}\mathcal{R}.
\end{equation}
Here $D_{OL}$ is the observer--lens distance, $\mathcal{R}$ is the flux ratio between the first relativistic image and the sum of the remaining images, and $r_m$ is the corresponding magnitude difference.

Expanding these observables to first order in $\beta$ gives
\begin{equation}
\begin{aligned}
\theta_\infty&=\frac{3\sqrt{3}M}{D_{OL}}\left[1-\frac{3\beta}{2}S(y)\right]+\mathcal{O}(\beta^2),
\\
r_m&=\frac{5\pi}{\ln 10}\left[1-\beta\mathcal{D}(y)\right]+\mathcal{O}(\beta^2).
\end{aligned}
\label{eq:thetarm2}
\end{equation}
The sign structure is immediate: for $\beta>0$, $\theta_\infty$ decreases (through $S>0$) and the magnitude difference $r_m$ also decreases (through $\mathcal{D}>0$), meaning a smaller angular shadow scale and weaker relative brightness contrast between images. For $\beta<0$, both trends reverse.

Eliminating $(\Omega_{\rm ph},\lambda_{\rm ph})$ in favor of $(\theta_\infty,\mathcal{R})$ gives the Stefanov-type map
\begin{equation}
\begin{aligned}
\operatorname{Re}\omega_{\ell n}&=\frac{Lc}{D_{OL}\theta_\infty}+\mathcal{O}(L^{-1}),
\\
\operatorname{Im}\omega_{\ell n}&=-\left(n+\frac12\right)\frac{c\ln\mathcal{R}}{2\pi D_{OL}\theta_\infty}+\mathcal{O}(L^{-1}).
\end{aligned}
\label{eq:Stefanov2}
\end{equation}
In Eq.~\eqref{eq:Stefanov2}, the factor $c$ is written explicitly to facilitate direct comparison with observational units; elsewhere in the analytic derivations we use $c=1$.

\section{Two limiting regimes and observational comments}

The core-size ratio $y=3M/\lambda$ separates two useful regimes:
Physically, $y$ compares the characteristic null-orbit scale ($r\sim3M$ in the Schwarzschild limit) with the radial extent of the scalarized core set by $\lambda$. Thus, $y\lesssim1$ means that the modified core reaches the photon-region scales and can leave comparatively strong imprints on shadow/ringdown/lensing observables, whereas $y\gg1$ means the core is well inside the photon sphere and its impact on external observables is parametrically suppressed. The intermediate regime $y=\mathcal{O}(1)$ is therefore the most informative region for disentangling core-size effects from coupling-amplitude effects in multi-channel analyses.
\begin{itemize}
\item \textbf{Wide-core limit} ($y\ll1$, i.e. $\lambda\gg M$):
\begin{equation}
\begin{aligned}
S(y)&=\frac{\pi}{2y}-\frac{2}{3}y^2+\mathcal{O}(y^4),\\
\mathcal{F}_\lambda(y)&=\frac{3\pi}{4y}-y^2+\mathcal{O}(y^4),
\\
\mathcal{D}(y)&=2y^4+\mathcal{O}(y^6).
\end{aligned}
\end{equation}
\end{itemize}
Accordingly, $\Omega_{\rm ph}$ and $\lambda_{\rm ph}$ acquire comparable leading corrections, while their ratio remains nearly unchanged; as a result, $\bar a$, $r_m$, and $\mathcal{Q}_n$ experience only mild modifications.
\begin{itemize}
\item \textbf{Compact-core limit} ($y\gg1$, i.e. $\lambda\ll M$):
\begin{equation}
\begin{aligned}
S(y)&=\frac{2}{y^2}-\frac{4}{3y^4}+\mathcal{O}(y^{-6}),
\\
\mathcal{F}_\lambda(y)&=\frac{1}{y^2}+\frac{4}{y^4}+\mathcal{O}(y^{-6}),
\\
\mathcal{D}(y)&=\frac{2}{y^2}-\frac{6}{y^4}+\mathcal{O}(y^{-6}).
\end{aligned}
\end{equation}
\end{itemize}
In this regime, corrections scale as $\lambda^2/M^2$ and reproduce the expected asymptotic RN-like behavior from Eq.~\eqref{eq:farRNlike}.

\begin{figure*}[t]
\centering
\includegraphics[width=0.99\linewidth]{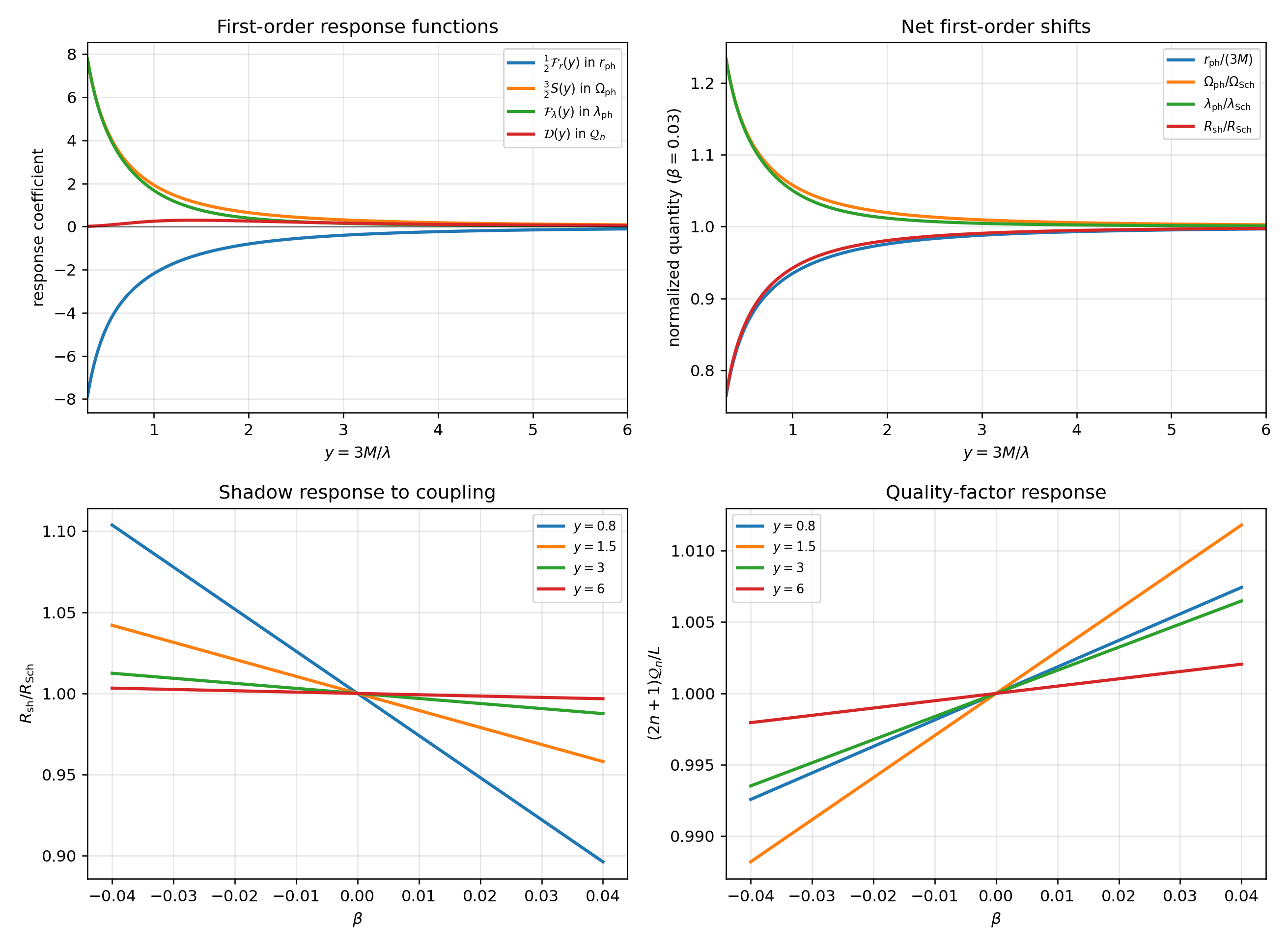}
\caption{Analytic first-order parameter response of key observables in the scalarized model. Top-left: response coefficients entering $r_{\rm ph}$, $\Omega_{\rm ph}$, $\lambda_{\rm ph}$, and $\mathcal{Q}_n$. Top-right: normalized shifts at representative coupling $\beta=0.03$. Bottom-left: linear dependence of $R_{\rm sh}/R_{\rm Sch}$ on $\beta$ for several $y$ values. Bottom-right: normalized quality-factor correction $(2n+1)\mathcal{Q}_n/L=1+\beta\mathcal{D}(y)$.}
\label{fig:parameter-sensitivity-main2}
\end{figure*}

Figure~\ref{fig:parameter-sensitivity-main2} provides a compact summary of these first-order responses, including the sign and magnitude of each coefficient as a function of $y$, the associated normalized shifts at a representative coupling, and the near-linear dependence on $\beta$ for selected core sizes. This visualization clarifies which observables are most sensitive in each region of parameter space.

Taken together, these limiting regimes isolate parameter degeneracies: the shadow size is primarily sensitive to $\beta S(y)$, whereas damping-ratio observables track $\beta\mathcal{D}(y)$. Consequently, a combined ringdown+shadow+lensing analysis can separate $(\beta,\lambda/M)$ more efficiently than any single observational channel.

It is important to delineate the domain of validity of the present treatment. The analytic formulas are derived within a weak-hair expansion and are therefore most reliable when $|\beta S(3M/\lambda)|\ll1$. Moreover, the ringdown and GBF correspondences are implemented at leading eikonal order, so quantitative accuracy deteriorates for low multipoles, high overtones, or in regimes where $\mathcal{O}(L^{-1})$ spin-dependent terms are non-negligible. In addition, the GBF model is constructed from the geodesic/QNM correspondence and does not replace full frequency-domain integration of the perturbation equations in precision applications. These limitations do not alter the qualitative trends derived above, but they define the regime in which the compact formulas should be interpreted as controlled approximations rather than exact predictions.

\section{Conclusion}

Working within the shift-symmetric, parity-preserving beyond-Horndeski scalar-tensor framework and its static, spherically symmetric scalarized black-hole solution introduced in Ref.~\cite{BakopoulosEtAl2024}, we developed an analytic first-order eikonal construction that links photon-sphere dynamics, quasinormal frequencies, shadows, strong lensing, and GBFs.

The core results are closed expressions for $r_{\rm ph}$, $\Omega_{\rm ph}$, and $\lambda_{\rm ph}$ in terms of two dimensionless combinations: the weak-hair strength $\beta=\eta q^4$ and the scale ratio $y=3M/\lambda$. These immediately generate the ringdown spectrum, shadow radius, strong-deflection coefficients, a compact quality-factor correction $\mathcal{D}(y)=2y^4/(1+y^2)^3$, and a closed GBF expression.

Beyond reproducing the standard eikonal correspondence structure, the new limiting analysis shows a practical distinction: for wide cores ($\lambda\gg M$), frequency and damping shifts track each other and ratio-type observables stay nearly stable, while for compact cores ($\lambda\ll M$), all corrections decay as $\lambda^2/M^2$. The GBF comparison at $\ell=3,4$ is consistent with this picture: changing $y$ primarily shifts the center of the transmission transition, whereas increasing $\ell$ mainly sharpens it.
Across all channels, the dominant qualitative pattern is therefore consistent: $\beta>0$ tends to increase characteristic frequencies ($\Omega_{\rm ph}$, $\operatorname{Re}\omega$), decrease characteristic angular scales ($R_{\rm sh}$, $\theta_\infty$), and strengthen damping ($|\operatorname{Im}\omega|$), with the largest fractional effects at intermediate-to-wide cores.

Several natural extensions follow from the present analysis, including the implementation of higher-order WKB/Pad\'e schemes \cite{MatyjasekOpala2017,KonoplyaZhidenkoZinhailo2019,Konoplya:2026fqh}, systematic comparisons with full numerical perturbation calculations beyond eikonal order, and confrontation of the analytic expressions with representative multi-messenger constraints.

\section*{Declaration of Competing Interest}
The authors declare that they have no known competing financial interests or personal relationships that could have appeared to influence the work reported in this paper.

\begin{acknowledgments}
B. C. L. is grateful to the Excellence project FoS UHK 2205/2025-2026 for the financial support.
\end{acknowledgments}

\section*{Data Availability}
No data was used for the research described in the article.

\bibliographystyle{apsrev4-2}
\bibliography{references}

\end{document}